\documentclass[article]{aa} 

\usepackage{natbib}
\usepackage{graphicx}
\usepackage{txfonts}
   \title{Dynamic properties of bright points in an active region}
	\authorrunning{Keys et al.}
	\titlerunning{Magnetic Bright Points Around An Active Region}
   \author{P.~H.~Keys\inst{1}\inst{,2}, M.~Mathioudakis\inst{1} , D.~B.~Jess\inst{1} , D.~H.~Mackay\inst{3}
          \and
          F.~P.~Keenan\inst{1} 
         }

   \institute{Astrophysics Research Centre, School of Mathematics and Physics, Queen's University Belfast, Belfast, BT7 1NN, Northern Ireland, U.K.\\
    \and
    Solar Physics and Space Plasma Research Centre (SP$^2$RC), The University of Sheffield, Hicks Building, Hounsfield Road, Sheffield, S3 7RH, U.K.\\
     \and
     School of Mathematics and Statistics, University of St. Andrews, North Haugh, St. Andrews, Fife, KY16~9SS, U.K. 
}

\offprints{P. Keys, \email{p.keys@qub.ac.uk}}

   \date{Received November 2013; accepted May 2014}

\begin{document} 

  \abstract
   {Bright points (BPs) are small-scale, magnetic features ubiquitous across the 
solar surface. Previously, we have observed and noted their properties for quiet Sun regions. Here, we determine the 
dynamic properties of BPs using simultaneous quiet Sun and active region data.
}
   {The aim of this paper is to compare the properties of BPs in both active and quiet Sun regions and to determine 
any difference in the dynamics and general properties of BPs as a result of the varying magnetic 
activity within these two regions.
}
   {High spatial and temporal resolution G-band observations of active region AR11372 were obtained with the Rapid Oscillations in the Solar Atmosphere 
instrument at the Dunn Solar Telescope. Three subfields of varying polarity and magnetic flux density were 
selected with the aid of magnetograms obtained from the Helioseismic and Magnetic Imager on board the Solar Dynamics Observatory. Bright points within these subfields were subsequently tracked and analysed.
}
   {It is found that BPs within active regions display attenuated velocity distributions with an average horizontal velocity 
of $\sim$0.6~km\,s$^{-1}$, compared to the quiet region which had an average velocity of 0.9~km\,s$^{-1}$.
Active region BPs are also $\sim$21\% larger than quiet region BPs and have longer average lifetimes ($\sim$132~s) than their quiet region counterparts (88~s). No preferential flow directions are observed within the active region subfields. 
The diffusion index ($\gamma$) is estimated at $\sim$1.2 for the three regions.
}
   {We confirm that the dynamic properties of BPs arise predominately from convective motions. The presence of stronger field strengths within active regions is the likely reason behind the varying properties observed. We believe that larger 
amounts of magnetic flux will attenuate BP velocities by a combination of restricting motion within the intergranular lanes and by increasing the number of stagnation points produced by inhibited convection. Larger BPs are found in regions of higher magnetic flux density and we 
believe that lifetimes increase in active regions as the magnetic flux stabilises the BPs.
}


   \maketitle
%

\section{Introduction}
Bright points (BPs) are ubiquitous in the solar atmosphere. With field strengths of the order of a kilogauss, they are thought to be the foot-points of magnetic flux tubes that have formed as magnetic flux is advected from the centre of granules 
into inter-granular lanes \citep{Sten85, Solan93}. Reduced pressure within the flux tube and the hot bottom effect result in BPs appearing as localised intensity enhancements in the photosphere \citep{Spru76,Stein98,Uit06}. Away from disk centre, the inclination of the partially evacuated flux tube allows the hot wall of the 
adjacent granule to be observed.
Observations in the G-band (4305{\AA}) are often used as a proxy for studying BPs as they appear brighter due to the reduced abundance of the CH 
molecule at higher temperatures \citep{Stein01}.

The magnetic field enclosed within the boundary of a BP can act to guide waves in the solar atmosphere \citep{deWi09}. Recent 3-D MHD 
models \citep{vBall11, Asgari12} have investigated the effect of Alfv{\'{e}}n waves generated by the motion of BPs. In addition to being linked to Alfv{\'{e}}n wave generation, BPs have been employed to  
identify locations of photospheric vortices \citep{Bon08} by observing rotational elements in their motions as they move towards downdraughts. They have also been observed in simulated studies \citep{Vog05} with localised vortex flows found at the location of some simulated BPs \citep{Shel11}. 
Observational work has also shown the  propagation of waves through BPs \citep{Jess12a}, while other research suggests they are the source of spicule oscillations  \citep{Jess12b}. 

Several studies have focused on BP properties such as size, velocity distribution and peak intensity 
\citep{Lan02, San04, Croc10, Keys2011}.  In particular, \citet{Keys2011} investigated the transverse velocities of over 6000 BPs using G-band observations of a quiet Sun region
and numerical simulations. The observed velocities were as high as 7~km\,s$^{-1}$, average velocities  were 1~km\,s$^{-1}$, and the BP lifetime was around 90~s. This study was extended to the upper photosphere/lower 
chromosphere using the Ca~{\sc{ii}}~K core where delays were found in the motion of BPs between the two layers \citep{Keys2013} along with elevated velocities observed in Ca~{\sc{ii}}~K BPs. A subsequent, independent study of {\sc{SUNRISE}} data by \citet{Jaf13} found 
similar elevated velocities in Ca~{\sc{ii}}~H BPs.  

\citet{Ber98} investigated quiet Sun and network BPs and found mean velocities of 1.1~km\,s$^{-1}$ and 0.95~km\,s$^{-1}$, respectively. The results were obtained with a temporal resolution of 25~s and a spatial resolution of 0$''$.83. \citet{Cris12} studied both a 
decaying active region and a quiet Sun region with the Interferometric BIdimensional Spectrometer \citep[{\it{IBIS}};][]{Cav06} and found that the velocities follow a Gaussian distribution with values as high as 2~km\,s$^{-1}$, but did not highlight any differences between
BPs found in an active region compared to a quiet region. Likewise, \citet{Mos06} observed BPs in a sunspot moat and found a mean velocity of 1.11$\pm$0.7~km\,s$^{-1}$. They suggested that the velocity values are attenuated in an active region by 0.2~--~0.35~km\,s$^{-1}$ compared to the solar network.

Another important component related to BP dynamics is dispersion \citep{Ber96, Ber98, Mos06, Utz10, Abra11}. The dispersion of magnetic elements is thought to be induced by photospheric flows \citep{Hag99}. A scalar diffusion coefficient ($K$) and the diffusion index ($\gamma$) are frequently 
estimated to determine the efficiency and the mechanism of diffusion respectively. The diffusion index, $\gamma$, can be established by monitoring the squared displacement ($sd$) of each BP over each time-step ($\tau$) in the BPs lifetime and is given by the relation $sd \propto \tau^{\gamma}$. If $\gamma = 1$ the 
dispersion is defined as {\textit{normal diffusion}} (i.e. a random walk) whereas when $\gamma < 1$ the process is described as {\textit{sub-diffusive}} and when $\gamma > 1$ the process is described as {\textit{super-diffusive}}.

Previous studies which use BPs as tracers in establishing this diffusion have found varying values for $K$ and $\gamma$ and therefore varying diffusive regimes. \citet{Law93} found a sub-diffusive regime in photospheric images of an active region with sub-diffusive motions also observed in a study by \citet{Cad99}. However, 
\citet{Cad99} reported that the sub-diffusive motions were found for timescales of 0.3--22 minutes, while normal diffusion occurs in the timescale of 25--57 minutes. More recent studies of photospheric BPs \citep{Abra11, Chit12, Gianna13} provide evidence of motions which are super-diffusive. \citet{Jafar14} also found 
super-diffusive motions in Ca~{\sc{ii}}~H images of the lower chromosphere/upper photosphere. It has been suggested \citep{Hag99} that estimations of the diffusion index may be effected by the lifetimes of the magnetic elements used to calculate the value.

In this study we compare the dynamic properties of BPs such as size, lifetime, diffusivity and horizontal velocity distributions in simultaneously observed active and quiet Sun regions, using high spatial and temporal resolution observations of the photosphere.


\section{Observations \& Data Reduction}
\label{obs}
\begin{figure*}[t]
\centering
{\scalebox{0.35}{\includegraphics{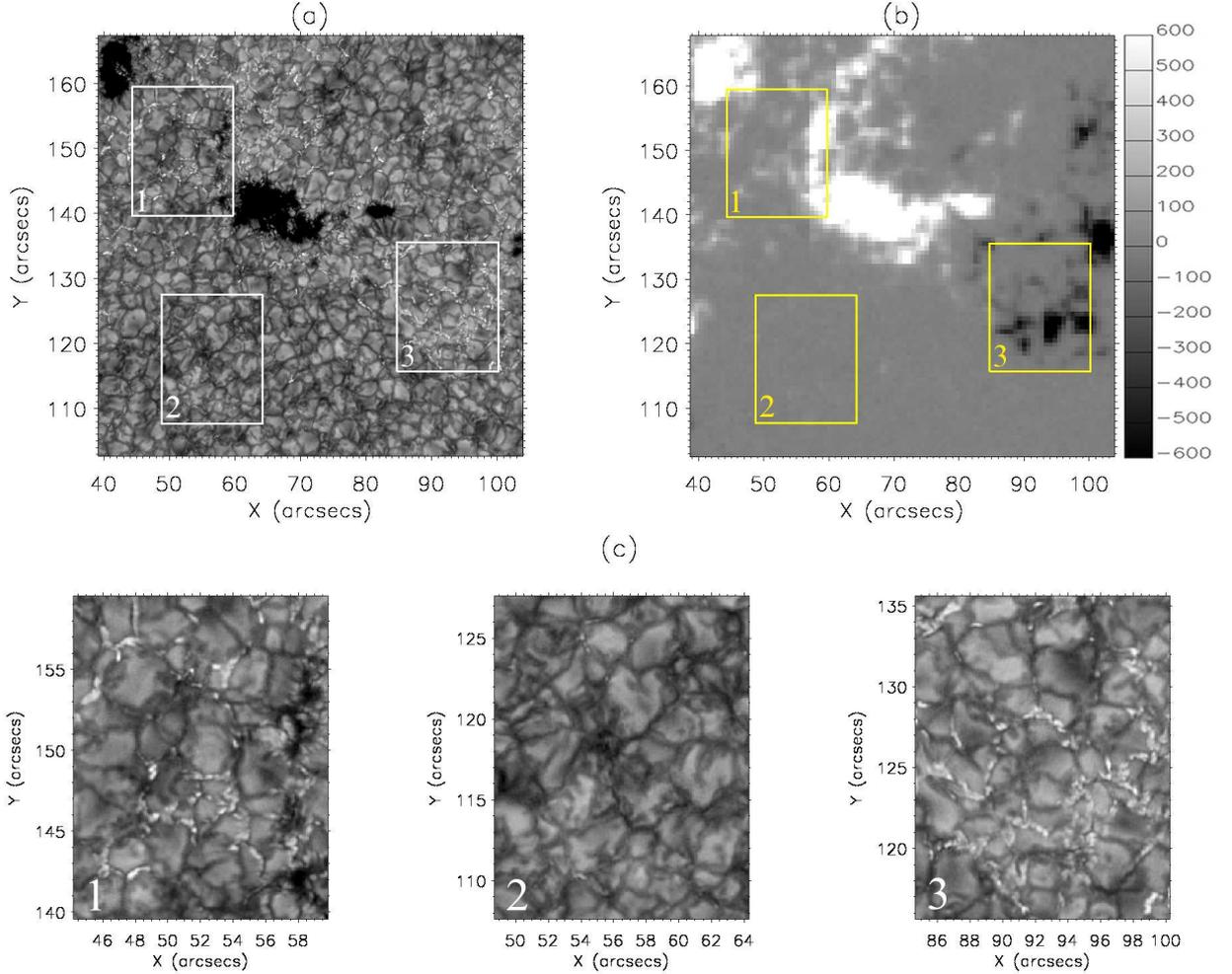}}}
\caption{{\textbf{(a)}} The full field-of-view of the ROSA G-band image of AR\,11372. {\textbf{(b)}} The corresponding 
HMI magnetogram. The {\textit{white}} boxes in {\textbf{(a)}} and the {\textit{yellow}} boxes in {\textbf{(b)}} show the three subfields used in the study which are expanded in {\textbf{(c)}}. The numbers in the boxes indicate which boxes in the full field-of-view correspond to their respective expanded views. 
A colour bar in {\textbf{(b)}} indicates the magnetic flux density (in gauss), saturated at $\pm$600~G to assist visualisation of the field 
complexities.}
\label{keys_fig_1}
\end{figure*}

Observations were obtained using the Rapid Oscillations in the Solar Atmosphere \citep[ROSA;][]{Jess10} instrument at the 76~cm 
Dunn Solar Telescope (DST) in New Mexico, USA. Data were acquired on 2011 December 10 under excellent seeing conditions that remained consistent for approximately 2~hours. The target was AR\,11372, located at N7.6W4.2 in the conventional heliographic co-ordinate system. 
For this study, a 9.2~{\AA} wide filter centred at 4305~{\AA} (G-band) was employed to study a 70$''\times$70$''$ 
field-of-view (Figure~\ref{keys_fig_1}) with a spatial scale of 0$''$.069~pixel$^{-1}$. Speckle reconstruction algorithms \citep{Wog08} were employed utilising 
bursts of 64 images for a single speckle reconstruction, resulting in a post reconstruction cadence of 2.112~s. Image destretching was performed using a 40$\times$40 grid \citep[equating to a $\approx$1.7$''$ separation between spatial samples;][]{Jess08}.

Line-of-sight magnetograms for the active region were obtained with the Helioseismic and Magnetic Imager \citep[HMI;][]{Schou12} on-board the Solar Dynamics 
Observatory \citep[SDO;][]{Pes12}. The instrument was employed with a cadence of 45~s and a two-pixel resolution of 1$''$.0. Additionally, one contextual 
HMI continuum image was obtained to aid the co-alignment of magnetograms with those from the lower solar 
atmosphere. The HMI data was processed with the standard {\verb+hmi_prep+} routine, including the removal of 
energetic particle hits. Subsequently, subfields were chosen from the processed data, 
with a central pointing similar to that of the ground-based image sequences. The HMI continuum image was used to define the absolute solar co-ordinates 
before the ground-based observations were subjected to cross-correlation techniques to provide sub-pixel co-alignment accuracy between the two imaging 
sequences. Following co-alignment, the maximum $x$- and $y$-displacements were both less than one tenth of a HMI pixel, or 0$''$.05 ($\approx$38~km). 
A sample of a co-aligned ROSA G-band image and a HMI magnetogram is shown in Figure~\ref{keys_fig_1}.

The spatial resolution of the reconstructed G-band images was estimated using methods outlined by \citet{Beck07} (Appendix A) and \citet{Pusch11}. A spatial resolution of around 0$''$.17 was established across a range of images in the 
dataset. This value was found to be roughly constant throughout and is close to the theoretical diffraction limit (0$''$.14) for the telescope at this wavelength.

\section{Data Analysis \& Results}
\label{analysis}
We selected three areas of equal dimensions (15.5$''\times$20$''$) from the G-band data for further analysis (Figure~\ref{keys_fig_1}\,(a)). 
These regions were selected based on a combination of the relative density of BPs in the region, and their position with respect to the polarity of the active region as determined from co-aligned HMI images (Figure~\ref{keys_fig_1}\,(b)).
One of the regions lies between two pores of the same polarity (average $\Phi$ = 169~G; average $| \Phi |$ = 169~G; Figure~\ref{keys_fig_1}\,(c): 1), the second in a relatively quiet region in the field-of-view (average $\Phi$ = 3~G; average $| \Phi |$ = 7~G; Figure~\ref{keys_fig_1}\,(c): 2) and the third in a region of opposite polarity to the first 
region, with reduced magnetic flux (average $\Phi$ = -101~G; average $| \Phi |$ = 107~G; Figure~\ref{keys_fig_1}\,(c): 3). It must be stressed that the dimensions of all three regions were kept equal so that a meaningful comparison
can be carried out. All results are summarised in Table~\ref{keys_table_1}.

Once the three regions were selected, the BPs in each were tracked using the algorithm employed by \citet{Keys2011}. Some of the most important aspects of this code will be discussed here, but for a full discussion 
see \citet{Croc09} and for improvements and additions implemented see \citet{Croc10} and \citet{Keys2011}.

The code has two main procedures; detection and tracking. The detection part of the code maps out the intergranular lanes and a binary map is produced of both the granules and the BPs. A compass search is then 
combined with intensity thresholding to remove granules from the resulting binary map. Because of the sharp turning points present at BP boundaries, this method allows an accurate determination of the boundaries. To avoid spurious detections 
of noise, the minimum threshold size for a BP to be detected is set at 4 pixels ($\approx$10\,000~km$^2$). \citet{Croc10} demonstrated that this is a suitable lower limit, with BP areas found to follow a log-normal distribution with a peak significantly 
above this limit. To avoid the detection of fragmenting granules, the upper limit for BP area is set at 200 pixels ($\approx$500\,000~km$^2$). This limit is sufficiently high to avoid excluding larger BP chains.

\begin{table*}[t!]
\centering    
\caption{Summary of BP characteristics in AR\,11372.}           
\label{keys_table_1}                                      
\scalebox{1.0}{
\begin{tabular}{l c c c}       
\hline\hline
{\textbf{Region}} & {\textbf{1}} & {\textbf{2}} & {\textbf{3}} \\
\hline     
{\textbf{No. of BPs}} & 2755 & 851 & 2843 \\
{\textbf{Av. $\Phi$ (G)}} & 169 & 3 & -105\\
{\textbf{Av. $| \Phi |$ (G)}} & 169 & 7 & 107\\
{\textbf{Av. Lifetime (s)}} & 136$\pm$40 & 88$\pm$23 & 128$\pm$38 \\
{\textbf{Av. Area (km$^2$)}} & 29\,000$\pm$2{\,}600 & 24\,000$\pm$2{\,}100 & 28\,000$\pm$2{\,}500 \\
{\textbf{Av. Velocity (km\,s$^{-1}$)}} & 0.6$\pm$0.3 & 0.9$\pm$0.4 & 0.7$\pm$0.4 \\
{\textbf{Max. Velocity (km\,s$^{-1}$)}} & 5 & 7 & 5.3 \\
{\textbf{Diffusion Index ($\gamma$)}} & 1.23$\pm$0.22 & 1.21$\pm$0.25 & 1.19$\pm$0.24\\
{\textbf{Diffusion Coefficient ($K$; km$^2$\,s$^{-1}$)}} & 109$\pm$43 & 123$\pm$45 & 137$\pm$56 \\
\hline                                            
	\end{tabular}
	}
\end{table*}

Once possible BP features are detected, the resulting binary map of detected features is passed through to the tracking algorithm to assign each detected feature with a tracking number and to establish essential properties such as area and the BP barycentre. 
For each BP detected, the algorithm searches an area around the detected BP in subsequent frames to find associated features. If another object falls within the search area, then the two features are considered as the same object and treated as such. If nothing 
is seen in the search area in subsequent frames, the object is considered to no longer exist. For cases in which a feature is missed in a frame before being found again in the next frame, e.g. due to a slight drop in seeing conditions in the data, the binary for the 
feature prior to being lost and immediately after being discovered again is used to determine the location of the missed feature. This ensures that the feature is tracked across all frames. Important information related to each feature, such as the area and barycentre 
for each frame, is then recorded for future use. For this study we ignore any BP that is found to originate from or found to pass beyond the boundary of our three selected regions. This alleviates certain issues that may arise in such scenarios, particularly with regards to the lifetimes of these features.

\begin{figure*}[t]
\centering
{\scalebox{0.25}{\includegraphics{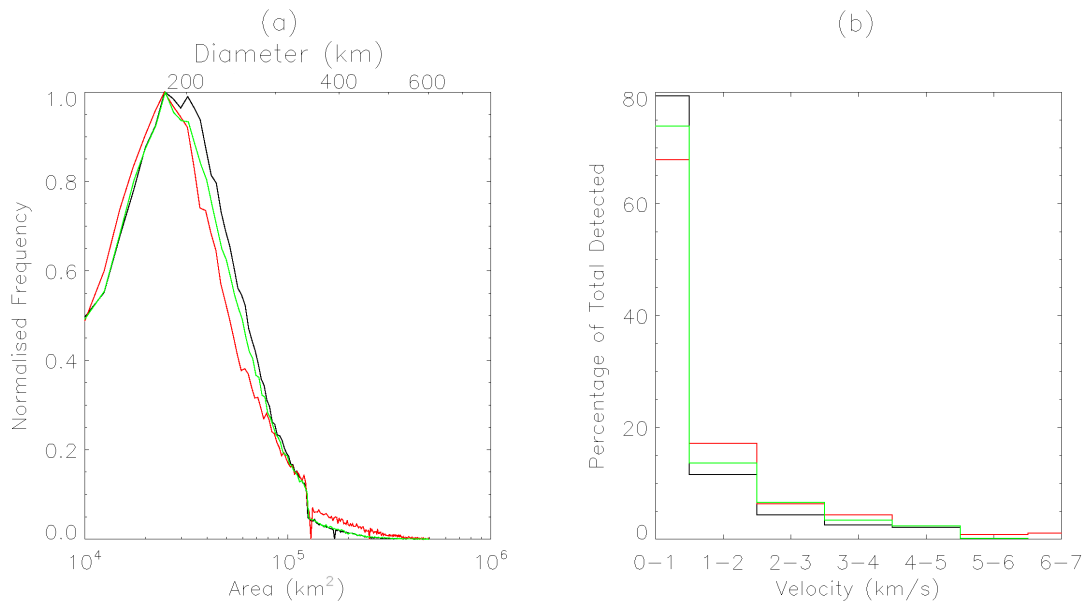}}}
\caption{{\textbf{(a)}} A semi-log graph of the area distributions for the BPs observed in region 1 ({\textit{black}}), region 2 
({\textit{red}}) and region 3 ({\textit{green}}). Regions with higher field strengths show slightly larger and broader peaks. Velocity histograms for the three subfields studied in AR\,11372 are shown in {\textbf{(b)}}. 
Again, {\textit{black}}, {\textit{red}}, and {\textit{green}} represent the distributions for region 1, 2, and 3, respectively. The quiet region (region 2) shows the presence of more elevated velocities 
than those found in the two more active regions.}  
\label{keys_fig_3}
\end{figure*}

The average horizontal velocity established for the first region in Figure~\ref{keys_fig_1} (hereafter region 1) was 0.6~km\,s$^{-1}$ 
from the 2755 BPs observed. For this region the highest velocity observed was 5~km\,s$^{-1}$, the 
average  BP lifetime was 136~s and with this the longest-living BP existed for $\sim$50~minutes. 
In the second region (hereafter region 2) we tracked 851 BPs which displayed remarkably similar characteristics to the quiet Sun results of \citet{Keys2011}. The average horizontal velocity for this region was found to be 0.9~km\,s$^{-1}$, with the highest velocity measured being 
$\sim$7~km\,s$^{-1}$. The corresponding average lifetime is 88~s and the longest-lived BP was observed for $\sim$15~minutes. 
Our third area was located between two regions of opposite polarity within AR\,11372 (hereafter region 3) and its BPs had characteristics similar to region 1. The average horizontal velocity 
of the 2843 BPs detected in this region was 0.7~km\,s$^{-1}$, maximum velocity 5.3~km\,s$^{-1}$, average BP lifetime 128~s and the maximum lifetime approximately 19.5~minutes. In Figure~\ref{keys_fig_3} we show the corresponding velocity distributions for each region. 

Our results highlight that BPs in active regions exist longer than their quiet Sun counterparts, with lifetimes of both region 1 (136~s) and region 3 (128~s) longer than region 2 (88~s).

The areas for the BPs were observed to follow a log-normal distribution (Figure~\ref{keys_fig_3}). This semi-log plot reveals peaks in region 1, 2, and 3 
at roughly 29\,000~km$^2$, 24\,000~km$^2$, and 28\,000~km$^2$, respectively. This gives a $\sim$21\% increase in size from the quiet to the more active regions. 
The size of a BP is often reported as the diameter that one would have if assumed to be a circle. With this, the peak diameter for the 
three subfields considered in this study are 190~km, 175~km, and 189~km for regions 1, 2, and 3, respectively. The number of BPs detected in both active subfields is 0.38 BPs per Mm$^2$, compared to only 0.09 
BPs per Mm$^2$ in the quiet region.

\begin{figure*}[t]
\centering
{\scalebox{0.35}{\includegraphics{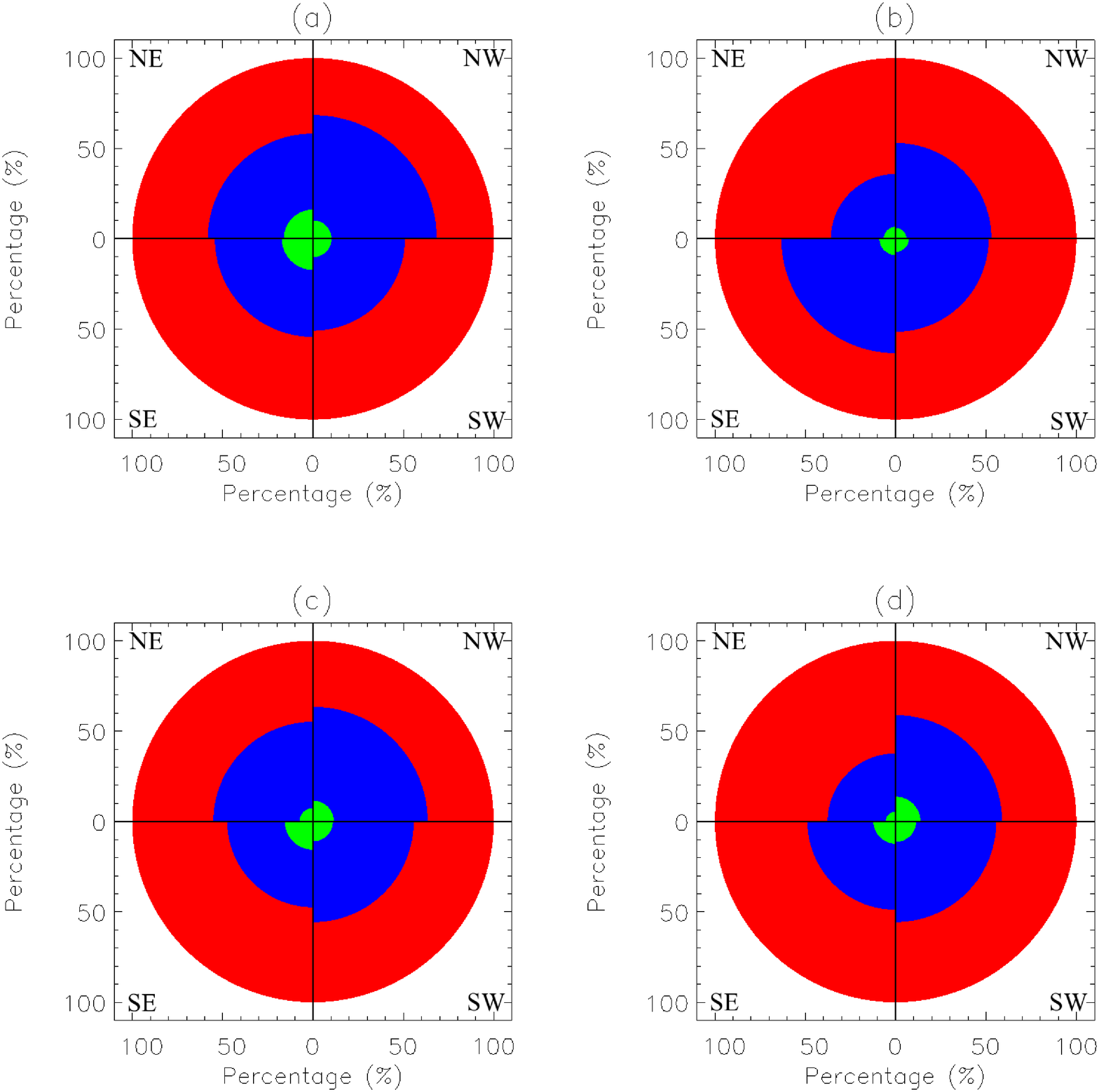}}}
\caption{Polar area diagrams showing flows in four quadrants in the three subfields. The percentage of BPs in 
a quadrant that move towards (\textit{blue}) and away from (\textit{red}) the 
reference point are represented alongside those BPs that experience no change throughout their lifetime 
(\textit{green}). The percentage for each flow direction in each individual quadrant is stacked as concentric arcs to aid the visualisation of flow directions within a quadrant. 
(a), (b), and (c) show the flows with reference to the centre of the subfield in region 1, 2, and 3 respectively. 
(d) shows the flows with respect to the neutral line found in region 3 with HMI LOS magnetograms.}
\label{keys_fig_4}
\end{figure*}

Given the ability of the algorithm to track large numbers of BPs, we decided to investigate if the BPs experience any preferential flow directions in their motions within each subfield. 
We examined the motions in relation to the centre-point of each of the three subfields and find no overall preferential flow direction (Figure~\ref{keys_fig_4}). In region 1, where the centre-point is between 2 pores of the same polarity,
43\% of the detected BPs move away from the centre of the dataset, about 48\% move towards the centre, while the remaining 9\% experience no change 
in direction. The average velocity of those BPs moving towards or away from the centre point is 1.1~km\,s$^{-1}$ and 1~km\,s$^{-1}$, respectively. Hence there is no real 
difference in the velocities of BPs moving towards or away from the centre of the subfield. The values are however slightly elevated in comparison to the average for the entire region  
(an overall average velocity of 0.6~km\,s$^{-1}$ was established for this dataset). This is expected due to the fact that a BP moving 
towards or away from the centre point will have some velocity component. By contrast, the overall distribution will also contain stationary BPs, which will lower the 
mean horizontal velocity component observed for the region as a whole. In region 2 it was observed that 50\% move away from the centre and 45\% move 
towards it, with the remaining 5\% experiencing no change in direction. Again, the average velocity for both moving towards and away from the centre point 
is slightly elevated at 1.5~km\,s$^{-1}$ and 1.4~km\,s$^{-1}$, respectively, when compared to the average velocity for the region. Within region 3 there is no preferential flow 
direction from the centre of the subfield, with 46\% moving away from the centre point, 46\% moving towards it and 8\% having no preferential direction. Velocities are similar to those found in region 1.

Additionally, we studied motions about the neutral line between two pores of opposite polarity in AR\,11372{\footnote{One of the pores is outside the field-of-view of ROSA.}}. The neutral line was found using LOS magnetograms of region 3 and was employed as a point of reference to establish the 
movements between these two pores. About 51\% of the BPs in the field-of-view move away from the LOS null-point and 41\% move towards it. The velocity values for  BPs moving away from the neutral line are 1.1~km\,s$^{-1}$, compared to 1.3~km\,s$^{-1}$ for those moving towards it. 
Figure~\ref{keys_fig_4} shows that there is some evidence of flows between quadrants in each subfield, however, more studies need to be carried out on various active region data sets to verify if these are real aspects of BP dynamics or anomalies.

We estimate the diffusion coefficient, $K$, and the diffusion index, $\gamma$, for the three regions. This is performed by monitoring the squared-displacement ($sd$) for each tracked BP at each time step ($\tau$) in their lifetime \citep{Cad99, Abra11,Rib11,Jafar14}. We establish 
$\gamma$ for each BP by plotting $sd$ as a function of $\tau$ on a log-log scale. We find $\gamma$ with the gradient of the plot as the two variables are related by the equation: 
\begin{equation}
sd(\tau) = C \tau^{\gamma},
\end{equation}
\noindent where $C$ is a constant of proportionality. As a value of $C$ can be evaluated from the log-log plot ($C = 10^{\gamma_{intercept}}$), we then establish the diffusion coefficient \citep{Mon75} with the equation:
\begin{equation}
K(\tau) = \frac{C \gamma}{4}\tau^{\gamma - 1}. 
\end{equation}

This method finds values for $\gamma$ in region 1, 2, and 3 as 1.23$\pm$0.22, 1.21$\pm$0.25, and 1.19$\pm$0.24, respectively. This suggests normal diffusion to slightly super-diffusive motions in the BPs in all three subfields. The corresponding 
diffusion coefficients for the three regions are 109$\pm$43~km$^2$\,s$^{-1}$, 123$\pm$45~km$^2$\,s$^{-1}$, and 137$\pm$56~km$^2$\,s$^{-1}$ respectively. The time ranges for the these estimates of the diffusion coefficients are $\sim$0.5--50~minutes, 
$\sim$0.5--15.5~minutes, and $\sim$0.5--19.5~minutes respectively.

\section{Discussion}
\label{discuss}

\subsection{Horizontal Velocity Properties and Lifetimes}

The histograms showing the velocity distributions for the three regions (Figure~\ref{keys_fig_3}) reveal, as do the average and maximum transverse velocity values, that BPs within areas of stronger 
magnetic field appear to have less dynamic behaviour. 
BPs with higher velocities ($>$2~km\,s$^{-1}$) have been linked 
to the production of magneto-sonic kink waves \citep{Choud93, deWi09}. Region 2 has the highest fraction of BPs with velocities $>$2~km\,s$^{-1}$, with 
about 15\% of the BPs studied having velocities greater than this value. This is higher than in the
two more-active regions, where the number of BPs with velocities greater 
than 2~km\,s$^{-1}$ in region 1 and 3 was estimated at 9\% and 12\%, respectively. It is expected that the higher field 
strengths within the active regions will inhibit convection by suppressing chaotic flows \citep{Cat03, Pal12}, which will then act to reduce the velocity distribution 
of BPs within regions of higher magnetic flux density, as is observed in the present study. This could explain the slightly lower horizontal velocities 
observed in region 1, as the HMI magnetograms show that the maximum flux density is higher in this region ($\approx$1700~G) 
compared to region 3 ($\approx$1100~G).

Our transverse velocities compare well with studies of BPs around active regions. Most of these \citep{Ber98,Mos06,Cris12} find mean 
horizontal velocities for BPs in active regions in the range 1~--~2~km\,s$^{-1}$, similar to those found here. It is also noteworthy that some 
observations \citep{Ber98, Mos06} find that BP velocities are attenuated in active regions compared to quiet Sun regions, as in the present work. 
Indeed \citet{Mos06} find that velocities in active regions are attenuated by about 0.2~--~0.35~km\,s$^{-1}$, while in our study the velocities between the quiet Sun region and the two active region subfields are also  attenuated by about 0.2~--~0.3~km\,s$^{-1}$. 

Also of note are the results with regard to region 2. The horizontal velocity value of 0.9~km\,s$^{-1}$ found for the 851 BPs studied is 
remarkably similar to the quiet Sun disk centre study of \citet{Keys2011}, who observed over 6000 BPs and determined an average velocity of 1~km\,s$^{-1}$. These 
results, coupled with the HMI magnetograms provides confidence that the subfield of region 2 within the active region data presented here is 
indeed a quiet region. This, therefore, leads to a meaningful comparison between the quiet region  and the two active region subfields. 
It should also be noted that the results presented here help improve our current understanding of the properties of BPs in active regions, as the 
dynamic properties are established simultaneously in quiet Sun and active region data.

With regards to BP lifetimes, the results for region 2 echo the 90~s 
average lifetime reported for the quiet Sun disk centre observations of \citet{Keys2011}.  \citet{Cris12} find similar lifetimes for quiet Sun BPs in their study. It
should be noted that \citet{Mos06} report a mean lifetime of 263$\pm$137~s for BPs observed in an active region which is longer than the value derived in the present study. However, the error bars are large in the case of \citet{Mos06}. Lifetimes of BPs can be difficult to ascertain exactly. The main issue arises in 
determining when the BP first appears and then disappears. Therefore the method 
employed for tracking the BP can have an impact on the lifetime. For example, methods that require human input in the identification process will tend to have larger lifetime estimates as the observer may actively look for associated features in subsequent frames and may introduce human error by actively trying to associate 
erroneous bright features to the BP being tracked. With fully-automated algorithms this is less of an issue but there needs to be some method for the algorithm to search for associated features or else the lifetimes will be underestimated. The algorithm used in this study looks 30~s ahead for associated features. If something is 
found, then it is treated as a continuation of the original feature. If nothing is observed within that time frame, then the feature is determined to have disappeared. 
We believe that the longer lifetimes displayed in the active region subfields in this study are the result of a higher magnetic flux density within these regions stabilising the BPs compared to quiet regions.

\subsection{Area Distributions and BP Numbers}

The area distributions shown in Figure~\ref{keys_fig_3} for the three regions follow a similar 													
log-normal distribution to those found in previous studies \citep{Ber95, Beck07, Croc10, Feng13, Keys2013}. \citet{Ber95} suggest that this distribution is the result of a fragmentation mechanism, similar to that found in sunspot umbrae \citep{Bog88}. As 
well as having a peak at a larger value, the width of the 
peak is also larger for the active regions compared to the quiet one. Such a broader distribution with a larger peak between active region and quiet Sun data is also observed by \citet{Rom12}. However,
a study by \citet{Feng13} suggests that there is little difference in both their active region and quiet Sun data with regards to the area distribution. Such differences can often be attributed to the resolution of the data 
\citep{Ber04, Utz09} and the feature tracking algorithm employed, specifically in this case, how the algorithm treats BP chains \citep{Pusch06}. 

In terms of BP diameters, a study of G-band BPs in an active region by \citet{Wie04} found a distribution of diameters between 100~--~300~km with the peak at 160~km. Similarly, \citet{Rom12} used IBIS to scan Fe~{\sc{i}} in both a bi-polar active region (7090~{\AA} scan) and a quiet Sun region (6302.5~{\AA} scan) and estimated 
the size distribution of BPs in both regions. In particular, they observed a 
peak in the active region distribution at 326~km and one in the quiet region at 181~km. However, the authors note that the high  value found in their active region distribution is likely the result of poor seeing conditions during the acquisition of the dataset and the lower spatial resolution of IBIS. \citet{Keys2013} find similar results on the effect of spatial resolution on BP size distributions. In their study, the 
photospheric level of a Na~{\sc{i}}\,D$_1$ IBIS scan observed a peak in BP size distributions at  a diameter of 255~km, whereas ROSA G-band images had a peak 
in the distribution at a diameter of 200~km. Given that the data in the current paper are from three subfields within the one field-of-view, our results are more 
readily comparable as all three subfields will be subjected to  the same seeing conditions over the duration of the observations.

The number of BPs found in the active region subfields (0.38 BPs per Mm$^2$) are again comparable to previous estimates 
of 0.3 BPs per Mm$^2$ \citep{San04} 
and 0.34 BPs per Mm$^2$ \citep{Rom12} in active regions.  However, the value for region 2 is quite low in comparison to other studies. For example, 0.22{\footnote{Found for images degraded to the seeing conditions of their active region data set.}} BPs per Mm$^2$ \citep{Rom12} and 0.19 BPs per Mm$^2$  
\citep{Keys2011}. This is likely a consequence of the smaller field-of-view used to obtain the results, or perhaps the region selected is particularly quiet and far away from 
a network boundary. It should also be noted that recent estimates by \citet{San10} indicate BP densities three times higher than those found previously, with a value of 0.97 BPs per Mm$^{2}$ reported. The authors 
attribute this difference to automated detection algorithms which they argue miss fainter objects. However, there is also an argument that over segmentation of longer BP chains is undesirable 
as it will increase the density of BPs reported. The next generation of solar telescopes and instruments will aid in improving these estimates of BP number densities and area distributions as they will allow 
a more accurate description of BP chains, and therefore, will allow us to determine if these chains can be segmented or not.

\subsection{Flow Directions and Diffusion}
We looked at the motions of all BPs tracked to determine if preferential flow directions were present in their motions. The overall numbers moving away/towards the centre point in each region, with the possible  exception of those about the location of the neutral line in region 3, show 
no particular preferential flow direction. We expanded on this by analysing the flow directions in each quadrant of the three regions about the centre-point and the neutral line found in region 3. The results in Figure~\ref{keys_fig_4} show that there are slightly elevated numbers flowing away or towards the 
reference point under investigation in some quadrants, for example, the polar area diagram for the motions with respect to the neutral line (Figure~\ref{keys_fig_4}\,(d)) suggests a greater number moving away from the neutral line in the NE quadrant (i.e. towards the pore near the centre of the data set). Given that 
only one active region was analysed in this study we cannot comment on whether such motions are a feature of the BPs dynamic properties with any degree of certainty. Also, with the lack of sufficient spatial resolution in regards to the magnetic field information for the BPs in this data set, we cannot 
formulate any theories or possible drivers that would explain these motions or if they are just random anomalies.

To aid in defining the reasons for the attenuation of velocities in region 1 and 3, we established the diffusion index ($\gamma$) and the diffusion coefficient ($K$) using the methods described by \citet{Abra11}. Our results suggest that the diffusion mechanism in all three regions is the same and shows that the 
BPs have normal diffusion (i.e. a random walk) to slightly super-diffusive motions. This is can be seen because we find values for $\gamma$ in region 1, 2, and 3 as 1.23$\pm$0.22, 1.21$\pm$0.25, and 1.19$\pm$0.24, respectively. 

The coefficient $K$ indicates the efficiency of the dispersion of the BPs as it measures the rate of increase in the dispersal area in units of time for all the BPs. Values for $K$ were found to be 109$\pm$43~km$^2$\,s$^{-1}$, 123$\pm$45~km$^2$\,s$^{-1}$, and 137$\pm$56~km$^2$\,s$^{-1}$ in region 1, 2, and 3 respectively. 
The values of $K$ that we establish in the three regions indicate that the BPs in region 3 experience the quickest rate of diffusion while those in region 1 experience the slowest rate of diffusion of the three subfields.  This may seem unusual, as one may expect regions with higher dynamic behaviour 
(such as region 2) to have a higher diffusion coefficient. However, the error bars on the value of $K$ estimated for region 3 are significantly larger than those for the other two regions. 

These results, to an extent, are similar to previous studies. \citet{Ber98} studied BPs in G-band and found values in the range 50--80~km$^2$\,s$^{-1}$. When the authors reconstructed the velocity field using correlation tracking techniques, 
they employed artificial passive corks which showed diffusion coefficients of 77, 176, and 285~km$^2$\,s$^{-1}$ in magnetic, network and quiet regions respectively. Likewise, \citet{Schri90} find values of 110 and 250~km$^2$\,s$^{-1}$ in the core of a plage region and 
surrounding quiet regions respectively. 

The results from region 2 in this study (i.e. the quiet region) are at odds with previous results. As highlighted in previous studies \citep{Hag99,Gianna13}, the temporal scales can effect the diffusion coefficient. \citet{Hag99} 
report diffusivities of 70--90~km$^2$\,s$^{-1}$ for timescales of less than three hours and values between 200--250~km$^2$\,s$^{-1}$ for timescales longer than eight hours. As the timescale that the values for region 2 are estimated from are considerably short, this may influence 
the diffusion values we find. Also, similar to the estimation of BP lifetimes, the method for tracking the BPs may effect the diffusion coefficients estimated. Taking the example of manual feature tracking again, by over-estimating the lifetime, the squared displacement values 
will be higher and there may be more L{\'{e}}vy flights in the BP paths which would increase the estimates of $\gamma$ and $K$. Furthermore, as diffusion can occur at varying spatial scales \citep{Manso11}, the total diffusion within the region under investigation is the resultant of the diffusivity at varying 
spatial scales (i.e. at granular and super-granular scales). This could lead to anomalies, however, we would only know for certain by expanding this study to numerous active region data sets.

The results with regards to the diffusion processes in these three regions do provide some important clues to the dynamic properties of BPs, however. The diffusion index established for the three regions shows that the same diffusion mechanism is found in both the 
quiet subfield and the active region subfields. This is important in that it shows that the motion of BPs is governed mainly by the buffeting motion of the granules within which these features are confined. This would suggest that the magnetic flux is passively tracing 
granular evolution. This process may explain the attenuation of BP velocities in active regions. In an active region where convection is inhibited by the magnetic field, there will be more stagnation points at the intersection of several granules. The motion of BPs in these 
stagnation points will be reduced and therefore will result in less dynamic behaviour and the attenuation of the velocities. Values for $K$ obtained previously at both quiet and active regions \citep{Schri90, Ber98} appear to support a greater number of stagnation 
points \citep{Gianna13} in active regions which would lead to reduced velocities compared to quiet regions.


\section{Conclusions}
\label{Conc}
The work presented here has studied the differences in the properties of BP found in both 
active regions and quiet Sun regions using high spatial and temporal resolution 
observations of the photosphere obtained with ROSA. Three subfields were considered in the G-band 
observations of an active region: an inter-pore region, a quiet region and a 
negative polarity region. These were chosen using HMI magnetograms and selected for their varying magnetic environments.

The results suggest that the transverse velocity component of BPs is reduced ($\sim$0.6~km\,s$^{-1}$) within the two active regions compared to 
BPs in quiet regions (0.9~km\,s$^{-1}$). This is believed to be the result of a combination of an increase in magnetic flux density within the intergranular lanes in active regions confining the BPs and the disruption of normal convection leading to an increase in the number of stagnation points. The net effect then is a reduction in the 
width of the velocity distributions of the BPs. BPs were observed to have longer lifetimes, on average, in the two high field strength regions ($\sim$132~s) compared to the quiet region (88~s). 
Again, this is attributed to the higher field strength within the active region subfields, which we believe produces BPs with greater stability. 
As expected, a higher number density of BPs is observed in the active regions (0.38 BPs per Mm$^2$), compared to the quiet region (0.09 BPs per Mm$^2$) which is also attributed to the 
higher magnetic flux density within the former. 

Similar to previous studies, a log-normal distribution of BP areas was observed for the active region subfields as well as for the quiet region. The peak of the 
distributions for the two active region subfields is both larger and broader than the quiet region subfield and suggests that BPs in active regions are $\sim$21\% larger than their counterparts in quiet regions. HMI images show that the average magnetic flux density in the quiet region (average $| \Phi |$ = 7~G) was much reduced in comparison to the other two regions under investigation (average $| \Phi | \approx $ 138~G).

The diffusion index for the three regions was also calculated and was found to be about 1.2 for the three regions. This would suggest that within the three subfields, the diffusion process is slightly super-diffusive for the tracked BPs. This has been shown in previous studies on diffusivity in the photosphere.
 As similar values are found across the three subfields, the results support the idea that the buffeting motion of the granules controls the dynamics of the BPs. In an active region, convective processes are inhibited by the higher magnetic flux density in the region. This results in a greater number of stagnation points within the active region 
which acts to attenuate the velocity characteristics of BPs found in the active region, which we observe in this study. More studies should be performed to determine the extent of this effect and to what extent the higher magnetic flux density within the lanes plays a role in reducing BP velocities.


\begin{acknowledgements} 
The authors wish to thank the anonymous referee whose comments enhance the work presented. 
PHK would like to thank Dr. Christian Beck for useful discussion of previous work. 
This work has been supported by the UK Science and Technology
Facilities Council (STFC). Observations were obtained
at the National Solar Observatory, operated by the Association
of Universities for Research in Astronomy, Inc. (AURA), under
cooperative agreement with the National Science Foundation. 
DBJ would like to thank the STFC for an Ernest Rutherford Fellowship.
We are also grateful for support sponsored by the Air Force Office of Scientific Research, 
Air Force Material Command, USAF under grant number FA8655-09-13085.
\end{acknowledgements}

\bibliographystyle{aa}

\begin{thebibliography}{}
\bibitem[Abramenko et al. (2011)]{Abra11} 
Abramenko, V.~I., Carbone, V., Yurchyshyn, V., et al., 2011, \apj, 743, 133
\bibitem[Asgari-Targhi \& van Ballegooijen (2012)]{Asgari12}
Asgari-Targhi, M., \& van Ballegooijen, A.~A., 2012, ApJ, 746, 81
\bibitem[Beck et al. (2007)]{Beck07} 
Beck, C., Bellot Rubio, L.~R., Schlichenmaier, R., S{\"{u}}tterlin, P., 2007, \aap, 472, 607 
\bibitem[Berger et al. (1995)]{Ber95} 
Berger, T.~E., Schrijver, C.~J., Shine, R.~A., et al., 1995, \apj, 454, 531 
\bibitem[Berger \& Title (1996)]{Ber96} 
Berger, T.~E., \& Title, A.~M., 1996, \apj, 463, 365
\bibitem[Berger et al. (1998)]{Ber98} 
Berger, T.~E., L{\"{o}}fdahl, M.~G., Shine, R.~A., \& Title, A.~M., 1998, \apj, 506, 439 
\bibitem[Berger et al. (2004)]{Ber04} 
Berger, T.~E., Rouppe van der Voort, L.~H.~M., L{\"{o}}fdahl, M.~G., et al., 2004, \aap, 428, 613 
\bibitem[Bogdan et al. (1988)]{Bog88} 
Bogdan, T.~J., Gilman, P.~A., Lerche, I., \& Howard, R., 1988, \apj, 327, 451 
\bibitem[Bonet et al.(2008)]{Bon08}
Bonet, J.~A., M{\'a}rquez, I., S{\'a}nchez Almeida, J., Cabello, I., \& Domingo, V., 2008, ApJL, 687, L131 
\bibitem[Cadavid et al. (1999)]{Cad99} 
Cadavid, A.~C., Lawrence, J.~K., \& Ruzmaikin, A.~A., 1999, \apj, 521, 844 
\bibitem[Cattaneo et al.(2003)]{Cat03} 
Cattaneo, F., Emonet, T., \& Weiss, N., 2003, \apj, 588, 1183 
\bibitem[Cavallini(2006)]{Cav06} 
Cavallini, F., 2006,  Sol. Phys., 236, 415 
\bibitem[Chitta et al. (2012)]{Chit12}
Chitta, L.~P., van Ballegooijen, A.~A., Rouppe van der Voort, L., DeLuca, E.~E., \& Kariyappa, R., 2012, \apj, 752, 48 
\bibitem[Choudhuri et al. (1993)]{Choud93}
Choudhuri, A.~R., Dikpati, M., \& Banerjee, D., 1993, ApJ, 413, 811
\bibitem[Criscuoli et al. (2012)]{Cris12} 
Criscuoli, S., Del Moro, D., Giorgi, F., et al., 2012, Memorie della Societa Astronomica Italiana Supplementi, 19, 93
\bibitem[Crockett et al. (2009)]{Croc09} 
Crockett, P.~J., Jess, D.~B., Mathioudakis, M., \& Keenan, F.~P., 2009, \mnras, 397, 1852  
\bibitem[Crockett et al. (2010)]{Croc10}
Crockett, P.~J., Mathioudakis, M., Jess, D.~B., Shelyag, S., Keenan, F.~P., \& Christian, D.~J., 2010, ApJL, 722, L188
\bibitem[de Wijn et al. (2009)]{deWi09}
de Wijn, A.~G., Stenflo, J.~O., Solanki, S.~K., \& Tsuenta, S., 2009, Space Sci Rev., 144, 275
\bibitem[Feng et al. (2013)]{Feng13} 
Feng, S., Deng, L., Yang, Y., \& Ji, K., 2013, \apss, 348, 17 
\bibitem[Giannattasio et al. (2013)]{Gianna13} 
Giannattasio, F., Del Moro, D., Berrilli, F., et al., 2013, \apjl, 770, L36 
\bibitem[Hagenaar et al. (1999)]{Hag99} 
Hagenaar, H.~J., Schrijver, C.~J., Title, A.~M., \& Shine, R.~A., 1999, \apj, 511, 932 
\bibitem[Jafarzadeh et al. (2013)]{Jaf13} 
Jafarzadeh, S., Solanki, S.~K., Feller, A., et al., 2013, {\aap}, 549, A116
\bibitem[Jafarzadeh et al. (2014)]{Jafar14} 
Jafarzadeh, S., Cameron, R.~H., Solanki, S.~K., et al., 2014, \aap, 563, A101
\bibitem[Jess et al. (2008)]{Jess08}
Jess, D. B., Mathioudakis, M., Crockett, P.~J., \& Keenan, F.~P., 2008, ApJL, 688, L119
\bibitem[Jess et al. (2010a)]{Jess10}
Jess, D. B., Mathioudakis, M., Christian, D.~J., Keenan, F.~P., Ryans, R.~S.~I., \& Crockett, P.~J., 2010a, Sol. Phys, 261, 363
\bibitem[Jess et al. (2012a)]{Jess12a}
Jess, D.~B., Shelyag, S., Mathioudakis, M., Keys, P.~H., Christian, D.~J., \& Keenan, F.~P., 2012a, ApJ, 746, 183
\bibitem[Jess et al. (2012b)]{Jess12b}
Jess, D.~B., Pascoe, D.~J., Christian, D.~J., Mathioudakis, M., Keys, P.~H., \& Keenan, F.~P., 2012b, ApJL, 744, L5
\bibitem[Keys et al.(2011)]{Keys2011}
Keys, P.~H., Mathoiudakis, M., Jess, D.~B., Shelyag, S., Crockett, P.~J., Christian, D.~J., \& Keenan, F.~P., 2011, ApJL, 740, L40
\bibitem[Keys et al.(2013)]{Keys2013}
Keys, P.~H., Mathioudakis, M., Jess, D.~B., Shelyag, S., Christian, D.~J., \& Keenan, F.~P., 2013, \mnras, 428, 3220 
\bibitem[Langhans et al. (2002)]{Lan02}
Langhans, K., Schmidt, W., \& Tritschler, A., 2002, A\&A, 394, 1069
\bibitem[Lawrence \& Schrijver (1993)]{Law93} 
Lawrence, J.~K., \& Schrijver, C.~J., 1993, \apj, 411, 402
\bibitem[Lemen et al.(2012)]{Lem12} 
Lemen, J.~R., Title, A.~M., Akin, D.~J., et al., 2012, \solphys, 275, 17 
\bibitem[Manso Sainz et al. (2011)]{Manso11} 
Manso Sainz, R., Mart{\'{\i}}nez Gonz{\'{a}}lez, M.~J., \& Asensio Ramos, A., 2011, \aap, 531, L9 
\bibitem[Monin \& Iaglom (1975)]{Mon75} 
Monin, A.~S., \& Iaglom, A.~M., 1975, Cambridge, Mass., MIT Press, 1975.~882
\bibitem[M{\"{o}}stl et al. (2006)]{Mos06} 
M{\"{o}}stl, C., Hanslmeier, A., Sobotka, M., Puschmann, K., \& Muthsam, H.~J., 2006, \solphys, 237, 13 
\bibitem[Pal \& Kumar(2012)]{Pal12} 
Pal, P., \& Kumar, K., 2012, European Physical Journal B, 85, 201 
\bibitem[Pesnell et al.(2012)]{Pes12} 
Pesnell, W.~D., Thompson, B.~J., \& Chamberlin, P.~C., 2012, \solphys, 275, 3 
\bibitem[Puschmann \& Wiehr(2006)]{Pusch06} 
Puschmann, K.~G., \& Wiehr, E., 2006, \aap, 445, 33
\bibitem[Puschmann \& Beck (2011)]{Pusch11} 
Puschmann, K.~G., \& Beck, C., 2011, \aap, 533, A21
\bibitem[Romano et al. (2012)]{Rom12} 
Romano, P., Berrilli, F., Criscuoli, S., et al., 2012, \solphys, 280, 407
\bibitem[Ribeiro et al. (2011)]{Rib11} 
Ribeiro, H.~V., Lenzi, E.~K., Mendes, R.~S., \& Santoro, P.~A., 2011, \physscr, 83, 045007 
\bibitem[S\'{a}nchez Almeida et al. (2004)]{San04}
S\'{a}nchez Almeida, J., M\'{a}rquez, I., Bonet, J.~A., Dom\'{i}nguez Cerde\~{n}a, I., \& Muller, R., 2004, ApJL, 609, L91
\bibitem[S{\'a}nchez Almeida et al. (2010)]{San10} 
S{\'a}nchez Almeida, J., Bonet, J.~A., Viticchi{\'e}, B., \& Del Moro, D., 2010, \apjl, 715, L26
\bibitem[Schou et al.(2012)]{Schou12} 
Schou, J., Scherrer, P.~H., Bush, R.~I., et al., 2012, \solphys, 275, 229 
\bibitem[Schrijver \& Martin(1990)]{Schri90} 
Schrijver, C.~J., \& Martin, S.~F., 1990, \solphys, 129, 95 
\bibitem[Solanki (1993)]{Solan93}
Solanki, S.~K., 1993, Space Sci. Rev., 63, 1
\bibitem[Shelyag et al. (2011)]{Shel11}
Shelyag, S., Keys, P.~H., Mathioudakis, M., \& Keenan, F.~P., 2011, A\&A, 526, A5
\bibitem[Spruit (1976)]{Spru76} 
Spruit, H.~C., 1976, {\solphys}, 50, 269
\bibitem[Steiner et al. (1998)]{Stein98} 
Steiner, O., Grossmann-Doerth, U., Knoelker, M., \& Schuessler, M., 1998, {\apj}, 495, 468 
\bibitem[Steiner et al. (2001)]{Stein01}
Steiner, O., Hauschildt, P.~H., \& Bruls, J., 2001, A\&A, 372, L13
\bibitem[Stenflo (1985)]{Sten85}
Stenflo, J.~O., 1985, Sol. Phys., 100, 189
\bibitem[Uitenbroek \& Tritschler (2006)]{Uit06} 
Uitenbroek, H., \& Tritschler, A., 2006, {\apj}, 639, 525 
\bibitem[Utz et al. (2009)]{Utz09} 
Utz, D., Hanslmeier, A., M{\"o}stl, C., et al., 2009, \aap, 498, 28
\bibitem[Utz et al. (2010)]{Utz10} 
Utz, D., Hanslmeier, A., Muller, R., et al., 2010, \aap, 511, A39
\bibitem[van Ballegooijen et al. (2011)]{vBall11}
Van Ballegooijen, A.~A., Asgari-Targui, M., Cranmer, S.~R., \& DeLuca, E.~E., 2011, ApJ, 736, 3
\bibitem[V{\"{o}}gler et al. (2005)]{Vog05} V{\"{o}}gler, A., Shelyag, S., Sch{\"{u}}ssler, M., et al., 2005, \aap, 429, 335 
\bibitem[Wiehr et al. (2004)]{Wie04} 
Wiehr, E., Bovelet, B., \& Hirzberger, J., 2004, \aap, 422, L63 
\bibitem[W\"{o}ger et al. (2008)]{Wog08}
W\"{o}ger, F., von der L\"{u}he, O., \& Reardon, K., 2008, {\aap}, 488, 375
\end{thebibliography}


\end{document}